\documentclass[useAMS,usenatbib]{mn2e}

\usepackage{natbib}
\usepackage{graphicx}
\usepackage{multirow}
\usepackage{subfig}

\title[Probing the Nuclear and Circumnuclear Activity of
NGC~1365 in the Infrared]{Probing the Nuclear and Circumnuclear Activity of
NGC~1365 in the Infrared\thanks{Herschel is an ESA space observatory with science instruments provided by European-led Principal Investigator consortia and with important participation from NASA.}}
\author[Almudena Alonso-Herrero et al.]
{\parbox{\textwidth}{A. Alonso-Herrero,$^{1,2}$\thanks{E-mail:
aalonso@ifca.unican.es} M. S\'anchez-Portal,$^{3}$  
  C. Ramos Almeida,$^{4,5}$ M. Pereira-Santaella,$^{6,7}$
P. Esquej,$^{6,1,8}$ S. Garc\'{\i}a-Burillo,$^{9}$ M. Castillo,$^{3}$
O. Gonz\'alez-Mart\'{\i}n,$^{4,5}$
N. Levenson,$^{10}$  E. Hatziminaoglou,$^{11}$ 
J. A. Acosta-Pulido,$^{4,5}$ J. I. Gonz\'alez-Serrano,$^{1}$
M. Povi\'c,$^{12}$ 
C. Packham,$^{13}$ and A. M. P\'erez-Garc\'{\i}a$^{4,5}$}
\vspace{0.4cm} \\
$^{1}$Instituto de F\'{\i}sica de Cantabria, CSIC-UC, Avenida de los
Castros s/n, 39005 Santander, Spain\\
$^{2}$Augusto Gonz\'alez Linares Senior Research Fellow\\
$^{3}$Herschel Science Centre, INSA/ESAC, E-28691 Villanueva de la
Ca\~nada, Madrid, Spain\\
$^{4}$Instituto de Astrof\'{\i}sica de Canarias, C/V\'{\i}a L\'actea
s/n, 38205 La Laguna, Tenerife, Spain\\
$^{5}$Departamento de Astrof\'{\i}ısica, Universidad de La Laguna,
38205 La Laguna, Tenerife, Spain\\
$^{6}$Centro de Astrobiolog\'{\i}a, INTA-CSIC, 28850 Madrid, Spain\\
$^{7}$Istituto di Astrofisica e Planetologia Spaziali, INAF-IAPS, Via Fosso
del Cavaliere 100, 00133 Rome, Italy\\
$^{8}$Departamento de F\'{\i}sica Moderna, Universidad de Cantabria,
Avenida de Los Castros s/n, 39005 
Santander, Spain\\
$^{9}$Observatorio Astron\'omico Nacional (OAN)-Observatorio de
Madrid, Alfonso XII 3, 28014 Madrid, Spain\\
$^{10}$Gemini Observatory, Casilla 603, La Serena, Chile\\
$^{11}$European Southern Observatory, Karl-Schwarzschild-Str. 2, 85748
Garching bei M\"unchen, Germany\\
$^{12}$Instituto de Astrof\'{\i}sica de Andaluc\'{\i}a, IAA-CSIC, C/ Glorieta de la Astronom\'{\i}a s/n, 18008 Granada, Spain\\
$^{13}$Astronomy Department, University of Florida, 211 Bryant Science
Center, PO Box 112055, Gainesville, FL 32611-2055, USA\\ 
}

\begin{document}

\date{Accepted --- . Received --- ; in original form --- }


\maketitle

\label{firstpage}

\begin{abstract}
We present new far-infrared ($70-500\,\mu$m) Herschel PACS and SPIRE imaging
observations as well as new mid-IR Gemini/T-ReCS 
imaging (8.7 and $18.3\,\mu$m) and spectroscopy 
of the inner Lindblad resonance (ILR)
region ($R<2.5\,$kpc)
of the spiral galaxy NGC~1365. We complemented these observations
with archival Spitzer imaging and spectral mapping observations. The
ILR region of NGC~1365  
contains a Seyfert 1.5 nucleus and a ring of star formation with an
approximate diameter of 2\,kpc. The
strong star formation activity in the ring is
resolved by the Herschel/PACS imaging data, 
as well as by the Spitzer $24\,\mu$m continuum emission, [Ne\,{\sc
  ii}]$12.81\,\mu$m line emission, and 6.2 and $11.3\,\mu$m
PAH emission. The AGN is
the brightest source in the central regions up to
$\lambda \sim 24\,\mu$m, but it becomes increasingly fainter in the
far-infrared when compared to the emission originating in the infrared
clusters (or groups of them) located
in the ring. We modelled the AGN unresolved infrared
emission with the {\sc clumpy} 
torus models and estimated that the AGN contributes only to a small  
fraction ($\sim 5\%$) of the infrared emission produced
in the inner $\sim 5\,$kpc.  
We  fitted the non-AGN $24-500\,\mu$m 
spectral energy distribution of the ILR region
and found that the dust temperatures and mass are similar to those of other
nuclear and circumnuclear starburst regions. Finally we showed that within the ILR region of
NGC~1365 most of the on-going star formation   activity is taking
place in dusty regions as probed by the $24\,\mu$m emission.  
\end{abstract}


\section{Introduction}

\begin{table*}

 \centering
 \begin{minipage}{100mm}
\caption{Summary of Herschel photometry observations}\label{tab:her_obs}

\begin{tabular}{lcccc}
\hline
Obsid & Instrument & Bands & Start time & Duration\\
& & (\micron) & (UTC) & (s)\\
\hline
1342201436 & SPIRE-P & 250, 350, 500 & 2010-07-14 20:20:45.0 & 999\\
1342222495 & PACS-P & 70, 160 & 2011-06-11 12:59:25.0 & 1217\\
1342222496 & PACS-P & 70, 160 & 2011-06-11 13:20:45.0 & 1217\\
1342222497 & PACS-P & 100, 160 & 2011-06-11 13:42:05.0 & 1217\\
1342222498 & PACS-P & 100, 160 & 2011-06-11 14:03:25.0 & 1217\\
\hline
\end{tabular}
\end{minipage}
\end{table*}

The fueling of the nuclear and circumnuclear activity of galaxies has
 been a topic of extensive discussion. Such activity not only
includes the accretion of matter onto a supermassive black hole with the
accompanying active galactic nucleus (AGN), but also the presence of 
intense nuclear and circumnuclear starbursts. Both types of activity
require gas to be transported from the host galaxy on physical scales
of a few kiloparsecs down  to less than one kiloparsec
for the nuclear starburst 
activity and even further in  for the nuclear
activity. Interactions, mergers,
and large-scale bars, among others, have been proposed as possible
mechanisms to transport gas from kiloparsec scales to the nuclear and
circumnuclear regions
\cite[see the review by][and references therein]{Jogee2006}.

In isolated galaxies with a large-scale bar the gas
is believed to flow inwards between corotation and the 
inner Lindblad resonance (ILR). Indirect evidence of this is the
presence of star formation
rings and hot spots near the ILR of barred galaxies \citep[see the 
review of][]{Buta1996}. The direction of the flows inside the ILR is 
generally outwards. This implies the existence of a gravity torque
barrier at this resonance. However, if there is a ``spatially-extended'' ILR
region, this translates into the existence of an inner ILR (IILR) and
outer ILR (OILR). Numerical simulations predict that in this case a 
gas ring can be formed in between these two resonances. Furthermore, in
the scenario of a double ILR the dynamical decoupling of an embedded
nuclear bar can drive the gas further in \citep{Shlosman1989,
  Hunt2008, GarciaBurillo2009}. The combined action of gravity torques
due to embedded structures (bars-within-bars)
and viscous torques could be a viable mechanism to drive the gas to
the inner few parsecs and feed the AGN \citep[][]{GarciaBurillo2005,
  Hopkins2011}.

NGC~1365 is a giant isolated barred galaxy at a distance of 18.6\,Mpc
\citep[][therefore $1\arcsec=90$\,pc]{Lindblad1999}.  
This galaxy hosts a Seyfert 1.5 nucleus \citep{Schulz1999}. 
\citet{Jungwiert1997} showed that in NGC~1365 there is also a nuclear bar
($R<10\arcsec$) embedded in the  the large-scale bar ($R\sim 100\arcsec$). 
\citet[][]{Lindblad1996} used hydrodynamical simulations to reproduce
the kinematics and the offset dust lanes along
the large-scale bar of this galaxy with an outer ILR at a radius
of $R_{\rm OILR}= 27\arcsec = 2.4\,$kpc. 
There is also an inner ILR at a
radius of $R_{\rm IILR} \sim 0.3\,$kpc. Henceforth we refer to the ILR
region as the region interior to the OILR of NGC~1365. As predicted by 
simulations, there is a ring of star formation inside the ILR region
of this galaxy. 

\begin{figure*}

   \subfloat{\includegraphics[width=0.46\textwidth]{figure1a.ps}}
   \subfloat{\includegraphics[width=0.51\textwidth]{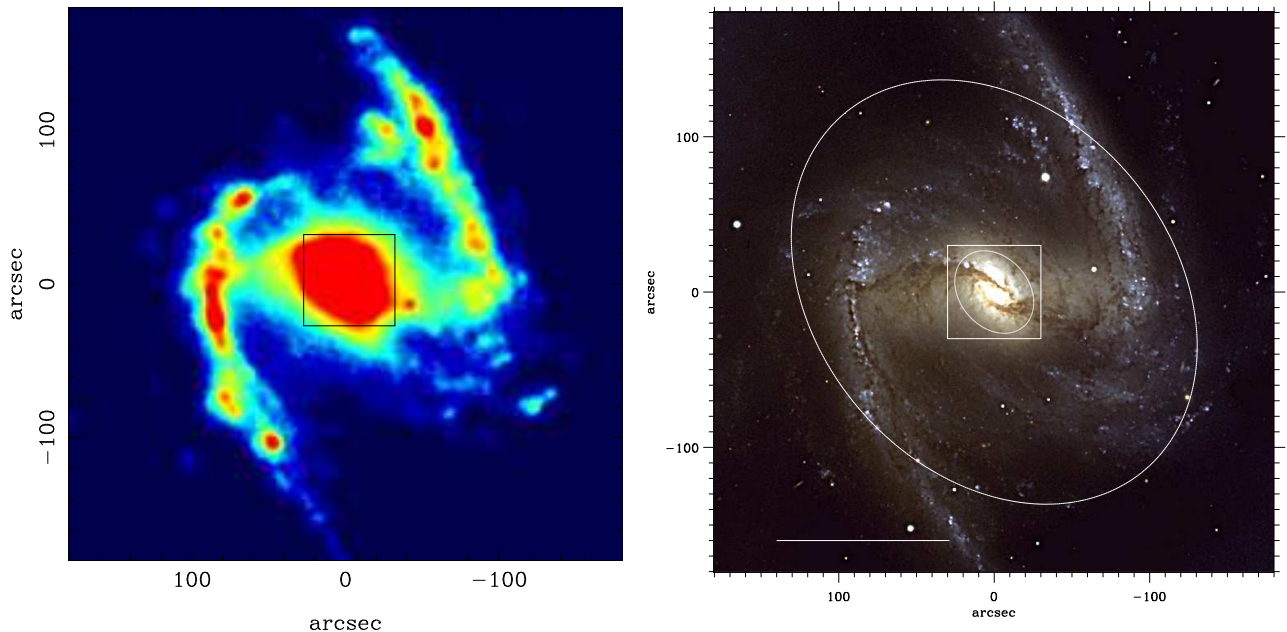}}


     \caption[PACS100_large]{Herschel/PACS $100\,\mu$m image (left
       panel) of NGC~1365. The image is shown in a square root 
      intensity scale. 
The displayed FoV is $360\arcsec \times
       360\arcsec$ and matches that of the optical image
       shown in the right panel. The latter reproduces the BVR image
       shown in figure~1 of \cite{Elmegreen2009} (reproduced by
       permission of the AAS), for an easy 
       comparison of the large-scale 
optical and far-IR morphologies. The square shows a
FoV$=60\arcsec\times60\arcsec$  
as in  Fig.~\ref{fig:IRcentralimages}. 
The small ellipse on the optical image 
represents the
       approximate size of the ILR region 
        studied in this work, whereas the large ellipse shows the
        corotation radius. The horizontal bar represents 10\,kpc. 
 For both images, orientation is north
        up, east to the left. } 
        \label{fig:PACS100_large}
    \end{figure*}

The star formation activity in the central regions of  
NGC~1365 has been
revealed by the presence of hot spots \citep{Sersic1965}, 
intense H$\alpha$ emission
\citep{Alloin1981, Kristen1997}, non-thermal radio continuum sources
associated with H\,{\sc ii} regions and 
supernova remnants \citep{Sandqvist1995, Forbes1998}, large amounts of 
molecular gas \citep{Sakamoto2007}, and point-like and diffuse
extended X-ray emission not associated with the AGN \citep{Wang2009}.
Moreover, there is evidence that a significant fraction of the circumnuclear 
star formation activity might be obscured based on the prominent dust
lane crossing the nuclear region (see Fig.~\ref{fig:PACS100_large},
right panel) and the bright and extended  mid-infrared (mid-IR)
emission in this region \citep{Telesco1993,
Galliano2005}.

In this paper we present new far-infrared (far-IR) imaging observations
of NGC~1365  performed with the ESA
Herschel Space Observatory \citep{Pilbratt2010} and new mid-IR    
imaging and spectroscopy 
obtained with the camera/spectrograph Thermal-Region Camera Spectrograph
\citep[T-ReCS;][]{Telesco1998} 
instrument on the Gemini-South telescope. The Herschel images were
obtained using the Photodetector Array Camera and 
Spectrometer \citep[PACS;][]{Poglitsch2010} and the Spectral and
Photometric Imaging REceiver 
\citep[SPIRE; ][]{Griffin2010} instruments. We also use archival
Spitzer data taken with the
Infrared Array Camera  \citep[IRAC;][]{Fazio2004}, the  Multi-Band
Imaging Photometer for Spitzer
\citep[MIPS; ][]{Rieke2004}, and the InfraRed Spectrograph 
\citep[IRS; ][]{Houck2004} instruments. 
Using IR observations to study the nuclear and circumnuclear 
activity in the ILR region of NGC~1365 
is crucial because the central region is crossed by a prominent
dust lane that obscures from our view in the optical a significant
fraction of emission produced there. 
This paper is organized as follows. We describe the observations in
Section~2. We analyze the AGN IR emission and 
the spatially resolved IR emission in Sections~3 and 4, respectively,
and we  summarize our conclusions in Section~5.

\begin{figure*}
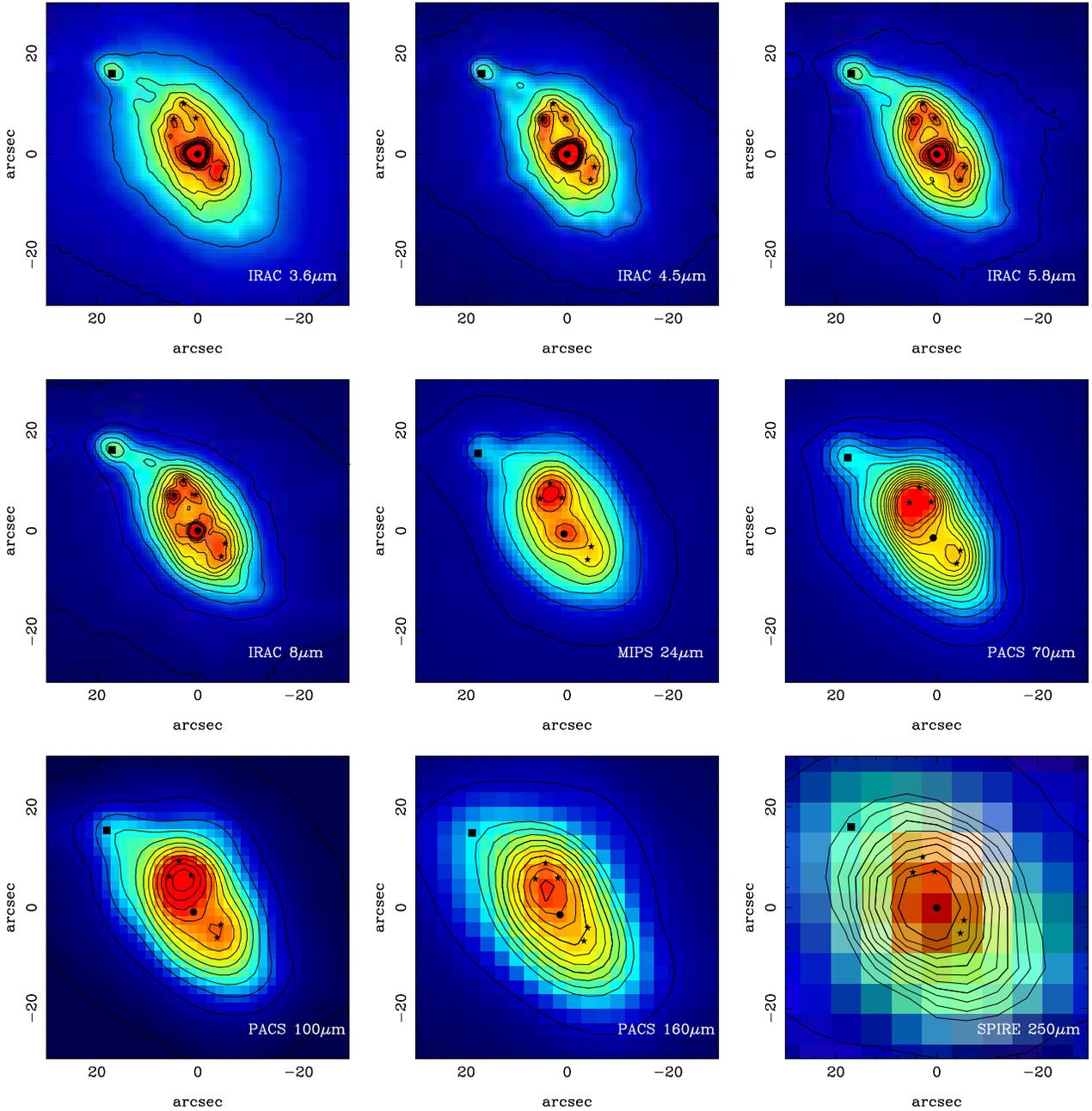

 
   \subfloat{\includegraphics[width=0.3\textwidth]{figure2a.ps}}
 \hspace{0.3cm}
 \subfloat{\includegraphics[width=0.3\textwidth]{figure2b.ps}}
 \hspace{0.3cm}
  \subfloat{\includegraphics[width=0.3\textwidth]{figure2c.ps}}
 
  \subfloat{\includegraphics[width=0.3\textwidth]{figure2d.ps}} 
 \hspace{0.3cm}
  \subfloat{\includegraphics[width=0.3\textwidth]{figure2e.ps}}
 \hspace{0.3cm}
  \subfloat{\includegraphics[width=0.3\textwidth]{figure2f.ps}}

  \subfloat{\includegraphics[width=0.3\textwidth]{figure2g.ps}}
 \hspace{0.3cm}
  \subfloat{\includegraphics[width=0.3\textwidth]{figure2h.ps}} 
 \hspace{0.3cm}
  \subfloat{\includegraphics[width=0.3\textwidth]{figure2i.ps}} 


     \caption[IRcentralimages]{IR view of the approximate extent of
       the ILR region  of NGC~1365. We show
        the Spitzer/IRAC images at  3.6, 4.5, 5.8, and $8\,\mu$m, 
        the Spitzer/MIPS image at $24\,\mu$m,  the
Herschel/PACS 70, 100, and $160\,\mu$m images and the Herschel/SPIRE
image at
$250\,\mu$m. We mark as filled
stars the positions of the brightest mid-IR clusters (M2, M3, M4,
M5, M6, see text for more details and also
Fig.~\ref{fig:closeups}) detected from the ground by   
\cite{Galliano2005}, the position of the AGN as a
filled dot, and the position of the L4 H$\alpha$ hot spot from
\cite{Alloin1981} as a filled square (see Table~\ref{tab:sources}).
Orientation is north up, 
east to the left. The images are shown in a square root intensity scale.}
        \label{fig:IRcentralimages}
    \end{figure*}

\section[]{Observations}

\subsection{Herschel/PACS and SPIRE imaging}
We obtained Herschel far-IR imaging observations of NGC~1365 
 using PACS at 70, 100, and $160\,\mu$m and SPIRE at 250, 350, and
 $500\,\mu$m. The data are part of the guaranteed time program
 entitled ``Herschel imaging photometry of nearby Seyfert galaxies: testing the
coexistence of AGN and starburst activity and the nature 
of the dusty torus''  (PI: M. S\'anchez-Portal). 

The PACS observations were carried out using the standard scan map
 mode that takes two concatenated scan line  
maps at  45$^{\circ}$ and 135$^{\circ}$ (in array coordinates), at a
speed of 20\,arcsec/sec, each one with 26 lines  
of $9\arcmin$ length and cross-scan step of $20\arcsec$. This mode
produces a rather homogeneous exposure map within a square region of
about  $7\arcmin \times 7\arcmin$. 
The set of maps were duplicated to observe through both the 70
\micron~(``blue'') and  100 \micron~(``green'') filters. Therefore the
galaxy was observed twice through the 160 \micron~(``red'') filter.  
With the SPIRE photometer we observed the three available bands 
simultaneously using the ``large map'' mode,  with two nearly
orthogonal scan maps (2 scan lines each), at a scan speed of
30\,arcsec/sec, and three repetitions.  The homogeneous exposure area
for scientific use is approximately
$8\arcmin \times 8\arcmin$. Table~\ref{tab:her_obs} gives the summary of
the observations. 

\begin{table*}
        
 \centering
 \begin{minipage}{100mm}
  \caption{Mid and far-IR aperture photometry of the central region of
    NGC~1365}\label{tab:photometry} 
  \begin{tabular}{lcccccc}
  \hline
  Instrument     &   $\lambda_{\rm c}$  & Pixel size & FWHM & 
 $f_\nu$ (r=15'') & $f_\nu$ (r=30'')\\
      &  $(\mu{\rm m})$ & (arcsec) & (arcsec) & (Jy) & (Jy)\\
 \hline
MIPS  & 24  & 2.45 & 5.9  & 7.2 & 8.7\\ 
PACS  & 70  & 1.4  & 5.6  & 85.5 & 102.7\\
PACS  & 100 & 1.7  & 6.8  & 110.1 & 141.7\\
PACS  & 160 & 2.8  & 11.3 & 87.4  & 123.7\\
SPIRE & 250 & 6.0  & 18.1 & 33.4  & 53.3 \\
SPIRE & 350 & 10.0 & 25.2 & 10.4  & 21.0 \\
SPIRE & 500 & 14.0 & 36.9 & --    & 5.6\\ 
\hline

\end{tabular}
Note.--- The reported values of the FWHM are the nominal values. The
errors in the aperture photometry are dominated by the photometric
calibration uncertainty of the instruments that is typically 10\%.
\end{minipage}

\end{table*}

We reduced the data with the Herschel Interactive
Processing Environment (HIPE) v8.0.1 and Scanamorphos  
\citep{Roussel2012} v15.  For the PACS instrument, we used HIPE and the
Calibration Database V32 to build Level 1 products. These included
detecting and flagging  bad pixels, converting the ADU readings to
flux units (Jy/pixel), and adding the pointing information. We did not
attempt to perform deglitching at this stage to prevent the bright AGN
nucleus to be affected by the MMT deglitching process. The final maps
were built from the Level 1 products using Scanamorphos, which
performs a baseline subtraction, correction of the striping effect due
to the scan process, removal of global and individual pixel drifts, and
finally the map assembly using all the nominal and cross-direction
scans.  For SPIRE we used the standard (small) map script and Calibration
Database v8.1. The processing included detection of
thermistor jumps in the time line, frame deglitching, low pass filter
correction, conversion of  readings to flux units (Jy/beam),
temperature drift and bolometer time response corrections, and addition
of pointing information. We built the final maps  using the
``na\"ive'' scan mapper task.  Colour  corrections (for PACS, see
\citealt{Poglitsch2010}; please refer to the \textit{Observer's Manual}
for  the SPIRE ones)  are small for blackbodies at the expected
temperatures (e.g., \citealt{PerezGarcia2001}) and  
have been neglected. More details on the processing of Herschel data
are given in S\'anchez-Portal et al. (in preparation).

Figure~\ref{fig:PACS100_large} shows the fully reduced PACS
$100\,\mu$m image of NGC~1365 together with the optical BVR image from
\cite{Elmegreen2009}. Fig.~\ref{fig:IRcentralimages} shows the
PACS images together with the SPIRE $250\,\mu$m with a field of view 
(FoV) covering the approximate extent of the ILR region of
NGC~1365 (see also Section~\ref{sec:alignment}). We performed 
aperture photometry on all the Herschel images using two different
radii, $r=15\arcsec$ and $r=30\arcsec$. The latter encompasses the
ILR region, whereas the former includes mostly the
ring of star formation. Table~\ref{tab:photometry} lists the measured 
flux densities.

\subsection{Gemini/T-ReCS imaging and spectroscopy}\label{sec:trecs}
We obtained 
mid-IR imaging of NGC~1365 using T-ReCS on the Gemini-South telescope on 
September 8 and 9, 2011 as part of 
proposal GS-2011B-Q-20 (PI: N. Levenson). 
We used the Si-2 ($\lambda_{\rm c}=8.74\,\mu$m) and
the Qa ($\lambda_{\rm c}=18.3\,\mu$m) filters and mid-IR standard
observation techniques. The plate scale of the T-ReCS imaging
observations is 0.089\arcsec/pixel with a FoV of $28.5\arcsec \times
21.4\arcsec$. The total integration
times (on-source) were 145\,s  and  521\,s in the Si-2 and Qa
filters, respectively. The Qa filter observations were split between
the two nights,  
whereas those in the Si-2 filter were done on the second night. We
observed standard stars immediately before or after the science
observations in the same filters, to both flux-calibrate the galaxy
observations and to estimate the unresolved nuclear emission (see
below). The
observations were diffraction limited, with a full width half
maximum (FWHM) of $0.34\arcsec$ in the
Si-2 filter and $0.55-0.58\arcsec$ in the Qa filter, as measured from 
the standard star observations. 
We reduced the imaging data using the CanariCam data reduction package
RedCan (Gonz\'alez-Mart\'{\i}n et al., in preparation). We refer the interested 
reader to this work and 
\cite{RamosAlmeida2011AGN} for  details.  Fig.~\ref{fig:trecs}
shows the fully reduced T-ReCS
Qa image resulting from the combination of the data taken during the
two observing nights.  The T-ReCS
$8.7\,\mu$m image (not shown here) shows a similar morphology. 

We also retrieved archival T-ReCS spectroscopic observations
 in the $N$-band ($\sim 8-13\,\mu$m) using a $0.35\arcsec$ slit width 
as part of proposal GS-2009B-Q-19  (PI: M. Pastoriza). The total
on-source integration time was 600\,s. We reduced the galaxy and
corresponding standard star observations using the RedCan package
following the steps described in \cite{AAH11AGN}. Finally we extracted the
nuclear spectrum as a point source. 

To estimate the nuclear unresolved emission from the mid-IR imaging data
we followed the point spread function (PSF) 
scaling technique  implemented by 
\cite{RamosAlmeida2009,RamosAlmeida2011AGN}. This unresolved emission which is
assumed to represent the
torus emission (see Section~\ref{sec:torusfit}). To do so, we scaled 
the observation of the corresponding standard star to the peak of the
nuclear emission of the galaxy in each of the two filters. This represents the
maximum contribution of the unresolved source (100\%), whereas 
the residual of the total emission minus the scaled PSF accounts for 
the extended emission. In both filters we found that a 90\% PSF
scaling provided a realistic estimate of the extended emission. 
The estimated  errors in the T-ReCS 
unresolved flux densities reported in Table~\ref{tab:AGNfluxes} 
are 15\% and 25\% in the Si-2 and Qa filters, respectively, and
account for both the flux
calibration and PSF subtraction uncertainties \citep[see][for more
details]{RamosAlmeida2009}. The unresolved $8.7\,\mu$m emission
computed this way is in good agreement with the flux density at the
same wavelength obtained from the T-ReCS nuclear spectrum.

\subsection{Archival Spitzer/IRAC and MIPS imaging}
We retrieved from the Spitzer archive imaging data of NGC~1365 observed with 
all four  IRAC  bands (3.6, 4.5, 5.8, and $8\,\mu$m)  and with MIPS at 
$24\,\mu$m  (Program ID: 3672, PI: J. Mazzarella). These
observations were part of The Great Observatories All-Sky LIRG Survey
\citep[GOALS,  
see][]{Armus2009}. We retrieved the basic calibrated data (BCD) from the
Spitzer archive. The BCD 
processing includes corrections for the instrumental response (pixel
response linearization, etc.), flagging of cosmic rays and saturated
pixels, dark and flat fielding corrections, and flux calibration based
on standard stars. 
We combined the BCD images into mosaics using the MOsaicker
and Point source EXtractor (MOPEX) software provided by the Spitzer 
Science Center using
the standard parameters. The final mosaics were repixeled to half of
the original pixel size of the images, that is, the IRAC mosaics have
$0.6\arcsec$/pixel, whereas the MIPS $24\,\mu$m mosaic has
$1.225\arcsec$/pixel. In Fig.~\ref{fig:IRcentralimages} we show 
the Spitzer images with a FoV covering the approximate extent of the 
ILR region, as done with the PACS images and the SPIRE $250\,\mu$m image. 
The angular resolutions of the IRAC images are between  1.7 and
$2\arcsec$  (FWHM)  and that of the MIPS
$24\,\mu$m is $5.9\arcsec$.

\subsection{Archival optical ground-based imaging}
We retrieved from the ESO archive optical images obtained with the WFI
instrument on the MPG/ESO 2.2m telescope using the narrow-band Halpha/7 filter
($\lambda_{\rm c}=6588.3$\AA, FWHM=70\AA) obtained as part of proposal 065.N-0076
(PI: F. Bresolin). 
We combined a total of 6 images, each of them with a
350\,s exposure. The plate scale of the images is
$0.238\arcsec$/pixel. The filter contains the H$\alpha$ and [N\,{\sc
  ii}]  emission lines plus
adjacent continuum. Since we use this image for morphological purposes
only, 
we did not attempt to either subtract the continuum or calibrate it
photometrically.  The positions of the H$\alpha$ hot spots identified
by \cite{Alloin1981} in the central region of NGC~1365 are displayed
in Fig.~\ref{fig:closeups} (upper panel). 

\begin{figure}

\hspace{0.5cm}
\resizebox{0.8\hsize}{!}{\rotatebox[]{0}{\includegraphics{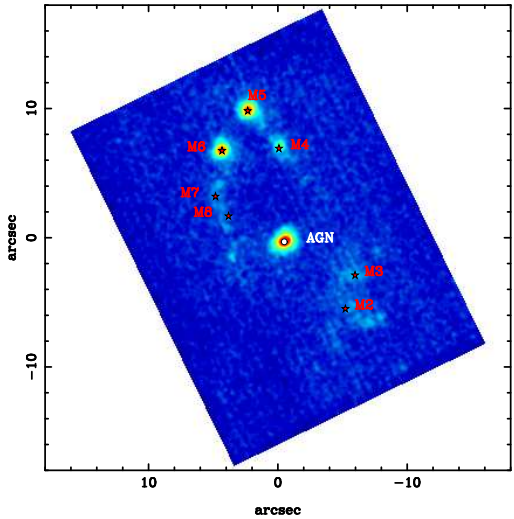}}}

     \caption[]{Gemini/T-ReCS image of the central region of NGC~1365
       in the Qa ($\lambda_{\rm c}=18.3\,\mu$m) filter. The image has
       been rotated to the usual orientation of north up, east to the
       left, and smoothed with a Gaussian function.   We mark the
         positions of the AGN (filled dot) as well as the  mid-IR  clusters
         (star symbols) identified by  \citet{Galliano2005}.}
        \label{fig:trecs}
    \end{figure}

  \begin{figure}
\hspace{0.5cm}
\resizebox{0.8\hsize}{!}{\rotatebox[]{0}{\includegraphics{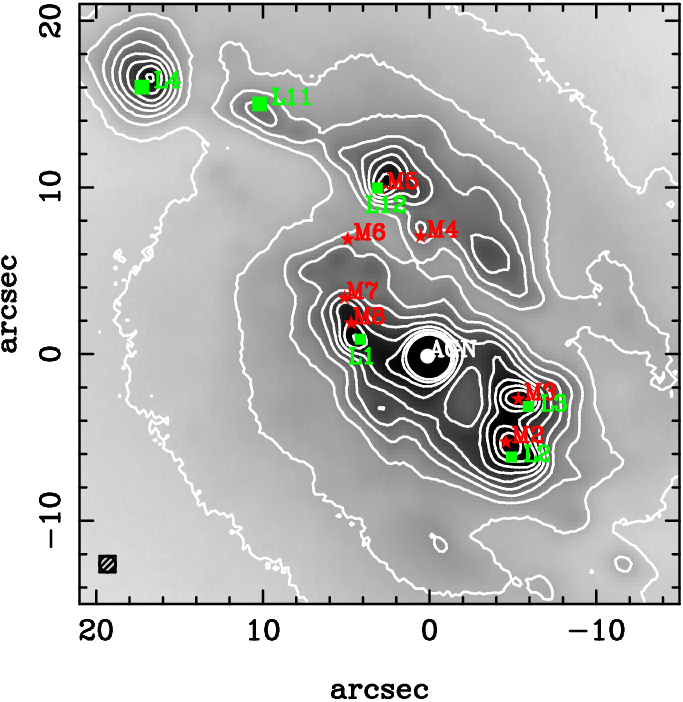}}}

\hspace{0.5cm}
\resizebox{0.8\hsize}{!}{\rotatebox[]{0}{\includegraphics{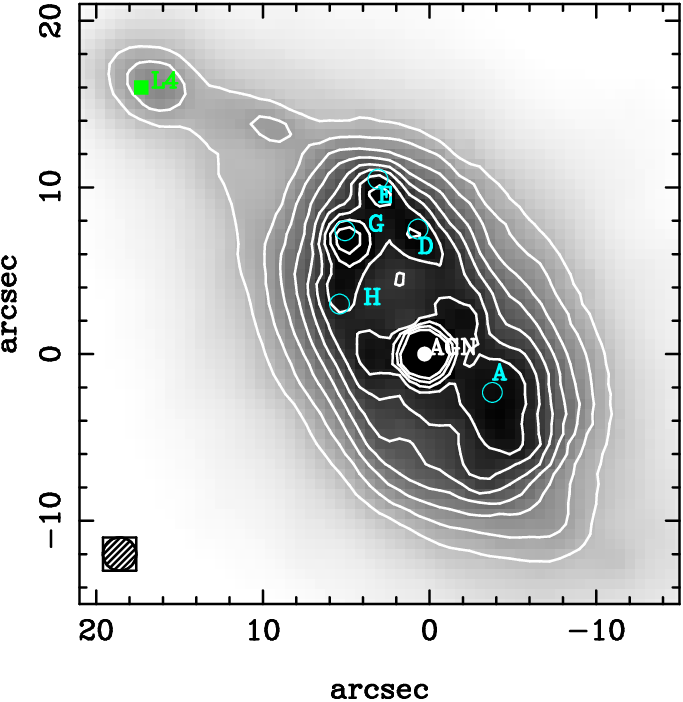}}}

\hspace{0.5cm}
\resizebox{0.8\hsize}{!}{\rotatebox[]{0}{\includegraphics{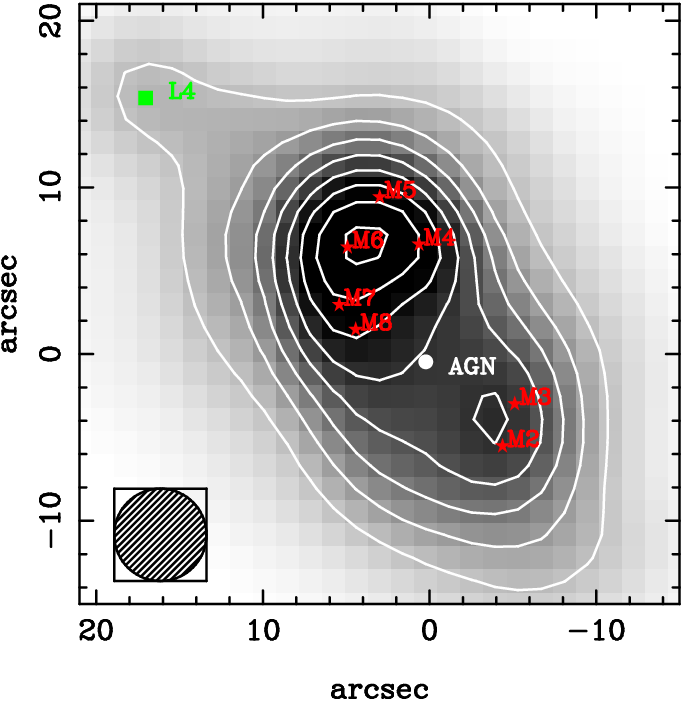}}}

     \caption[]{Close-ups of the inner $36\arcsec \times 36\arcsec$ region
       showing the 
location of the AGN (filled dot), the mid-IR clusters M2...M8
(filled star symbols) of \cite{Galliano2005}, the radio sources 
(open circles) of \cite{Sandqvist1995}, and the H$\alpha$ hot spots 
(filled squares) of
\cite{Alloin1981} and \cite{Kristen1997}. The upper panel is the
optical ESO Halpha/7 narrow-band image, the middle panel is the IRAC
$8\,\mu$m image, and the lower panel the PACS $70\,\mu$m image. The beam  of the 
images  is shown on the lower left corner of each image,
approximated as the FWHM of a Gaussian function.}
\label{fig:closeups}
    \end{figure}

\subsection{Alignment of the images}\label{sec:alignment}
The alignment of the Spitzer/IRAC and the Gemini/T-ReCS images 
is straightforward because
the AGN and the mid-IR clusters detected by
\cite{Galliano2005} are clearly identified in all these images. The AGN
is also bright in the optical image and therefore we used it as our 
reference. In
Fig.~\ref{fig:closeups} we present a close-up of the ILR region in the IRAC
$8\,\mu$m band and the optical narrow-band H$\alpha$ image. We
marked the positions of the bright (designated as M4, M5, and M6) and
faint mid-IR 
(designated as M2, M3, M7, and M8) clusters
and the
nucleus using the relative positions given by \cite{Galliano2005}. We
also indicated the positions of radio sources and H$\alpha$ sources
(see Section~\ref{sec:morphology_detailed}). Although
the Spitzer/MIPS $24\,\mu$m image has a poorer angular resolution
when compared to that of
IRAC, the AGN is still sufficiently bright at $24\,\mu$m that can be
distinguished  from  the mid-IR clusters.

The alignment of the Herschel/PACS images is not as simple because 
the AGN does not appear to be a bright source in the far-IR. 
We used  the astrometry information in the  Herschel
image headers and the optical position of the nucleus of NGC~1365
given by \cite{Sandqvist1995}:
RA(J2000)=$03^{\rm   h}33^{\rm m}36.37^{\rm s}$ and
Dec(J2000)=$-36^{\circ}08\arcmin25.5\arcsec$, for the initial
alignment.  
These coordinates placed the AGN to the southwest of the
bright source detected in the PACS images and in the MIPS $24\,\mu$m
image that appears to coincide with the region containing 
clusters M4, M5, and M6. 
Additionally we compared the morphologies of the 
PACS $70\,\mu$m and the MIPS $24\,\mu$m images as they have
similar angular resolutions (see Table~\ref{tab:photometry}).
We used the mid-IR source 
located $\sim 17\arcsec$ N, $16\arcsec$ E of the AGN 
identified in the IRAC and MIPS images
and also seen in the PACS $70\,\mu$m and $100\,\mu$m images 
for a finer alignment. 
This source appears to be 
coincident with the L4 H\,{\sc ii} region or H$\alpha$ hot spot  
(see Fig.~\ref{fig:closeups} and Table~\ref{tab:sources}) identified by
\cite{Alloin1981} and also seen in our archival H$\alpha$ image. The
PACS $160\,\mu$m image was aligned relative to the other PACS images
with the astrometry information in the headers. Finally  since the
bright mid-IR clusters cannot be resolved in the SPIRE
$250\,\mu$m image, we placed the center of image at the position of
the AGN. 
 We note that the 
alignment of the PACS images in Figs.~\ref{fig:IRcentralimages} and
\ref{fig:closeups} is only good to within 1 
pixel in each band.
Since the main goal of this work is to
study the processes giving rise to the IR emission within the ILR
region of
NGC~1365, in Fig.~\ref{fig:IRcentralimages} we show the aligned IR images 
with a FoV covering the approximate extent of this region. We do not
show the SPIRE 350 and $500\,\mu$m images because of the  small number
of pixels covering the ILR region of NGC~1365.

\begin{table}
        
 \centering
 \begin{minipage}{70mm}
  \caption{AGN emission}\label{tab:AGNfluxes}
  \begin{tabular}{lcc}
\hline
Wavelength & $f_\nu$ & Method \\
($\mu$m)   & (mJy)\\
\hline
8.7   & $203\pm30$ & Imaging (unresolved) \\ 
13.0    & $400\pm60$         & Spectroscopy \\
18.3 & $818\pm205$ & Imaging (unresolved)  \\
24    & $1255^{+783}_{-500}$ & BC fit \\
70    & $734^{+1482}_{-422}$  & BC fit\\
100   & $271^{+632}_{-163}$   & BC fit\\
160   & $<78$    & BC fit\\
\hline
\end{tabular}

References.---  The quoted uncertainties for the BC fit fluxes are the
$\pm 1\sigma$ uncertainties, as
discussed in Sections~\ref{sec:torusfit} and \ref{sec:farIRAGNemission}.
\end{minipage}
\end{table}

\subsection{Spitzer/IRS spectral mapping}
We retrieved from the Spitzer archive 
low spectral resolution ($R \sim 60-126$) 
observations (Program ID: 3269, PI: J. Gallimore) of NGC~1365 obtained with the 
spectral mapping capability of IRS. These observations were part of
the Spitzer study of the spectral energy distributions (SED) of the
$12\,\mu$m sample of active galaxies \citep{Gallimore2010}.
The observations were obtained with 
the Short-Low
(SL1; $7.5-14.3\,\mu$m  and SL2; $5.1-7.6\,\mu$m) 
and Long-Low (LL1; $19.9-39.9\,\mu$m and LL2; $13.9-21.3\,\mu$m) 
modules. The plate scales of the SL and LL modules are
$1.8\arcsec$/pixel and $5.1\arcsec$/pixel, respectively.

\begin{figure}
\resizebox{0.96\hsize}{!}{\rotatebox[]{-90}{\includegraphics{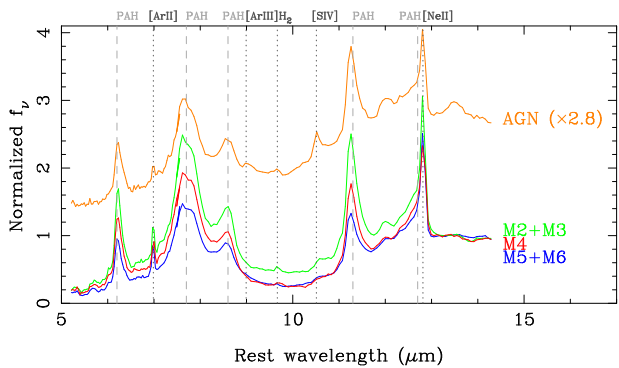}}}
\vspace{-1.5cm}
     \caption[]{Spitzer/IRS SL1+SL2 $5-15\,\mu$m spectra of selected
       regions (see 
       Table~\ref{tab:spectroscopy}) normalized at $13\,\mu$m. 
         The spectrum of the AGN has been shifted up for clarity. The most
       prominent spectral features are marked.}
        \label{fig:SLspectra}
    \end{figure}

\begin{table}
        
 \centering
  \caption{Summary of sources in the circumnuclear region}\label{tab:sources}
  \begin{tabular}{lccccccc}
\hline
Name & Spectral &Rel. Position & Ref. \\
  & Range & arcsec, arcsec & \\
\hline
M2, L2 & mid-IR, H$\alpha$ &  -4.7, -5.1 & 1, 2\\
M3, L3 & mid-IR, H$\alpha$ &  -5.4, -2.6 &1, 2, 3\\
M4, D  & mid-IR, radio     & 0.4, 7.1 & 1, 4\\
M5, E, L12 & mid-IR, radio, H$\alpha$ & 2.8, 10.0 & 1, 4, 3\\
M6, G & mid-IR, radio      & 4.8, 7.0 & 1, 4 \\
M7, H& mid-IR, radio & $^*$ & 1, 4\\
M8, L1 & mid-IR, H$\alpha$ & $^*$ & 1, 2\\
L11 & H$\alpha$ &10, 15 & 3\\
L4 & H$\alpha$ & 17, 16 & 3\\
A  & radio &-4.1, -2.4 & 4\\
\hline
\end{tabular}

The positions are relative to that of the AGN and correspond to those
of the first listed source. $^*$The positions shown in Figs.~3 and 4 are 
estimated from the mid-IR images of Galliano et al. (2005).
References.--- 1.  Galliano et al. (2005). 2. Kristen et
al. (1997). 3. Alloin et al. (1981). 4. Sandqvist et al. (1995). 
\end{table}

The data were processed using the Spitzer IRS pipeline.
The IRS data cubes were assembled using {\sc cubism} 
\citep[the CUbe Builder for IRS Spectra Maps,][]{Smith2007CUBISM} v1.7 from the
individual BCD  
spectral images obtained from the Spitzer archive. 
{\sc cubism} also provides error data cubes 
built by standard error propagation, using, for the input uncertainty, 
the BCD-level uncertainty estimates produced by the IRS pipeline  
from deviations of the fitted ramp slope fits for each pixel. 
We used these uncertainties to provide error
estimates for the extracted spectra, and the line and continuum
maps \citep[see][for full details]{Smith2007CUBISM} discussed in
the next two sections.

\subsubsection{Extraction of the 1D spectra}\label{sec:1Dspectra}


We used {\sc cubism} to extract 
spectra of regions of interest using small apertures 
taking advantage of the  angular resolution of the 
spectral mapping observations obtained with the SL1+SL2 modules
\citep[$\sim 4\arcsec$ 
FWHM, see][]{Pereira2010IRSmapping}. We used square or
rectangular apertures 
in the original orientation of the SL data
cubes with sizes of two or three pixels (see
Table~\ref{tab:spectroscopy} for the extraction apertures used
  for each region). The selected regions  include the  
AGN,  and the regions containing the M2+M3, M4, and M5+M6 mid-IR
clusters (see 
Fig.~\ref{fig:closeups} for the positions) identified by
\cite{Galliano2005}. We note that we did not attempt to
  apply a point source correction to the SL $3.7\arcsec \times 3.7\arcsec$
  spectrum of the AGN  because we are mostly interested in the extended
  features, that is, the polycyclic aromatic hydrocarbon (PAH) 
features and the [Ne\,{\sc ii}]$12.81\,\mu$m
  fine structure line. Fig.~\ref{fig:SLspectra} shows the SL1+SL2
  spectra of the selected regions  normalized at $13\,\mu$m. Finally we  
extracted the integrated spectrum of the region covered by the
observations, i.e. the central
  $27.8\arcsec \times 24.0\arcsec$.

We measured the fluxes and the equivalent widths (EW) of the
[Ne\,{\sc ii}]$12.81\,\mu$m emission 
line and the 6.2 and $11.3\,\mu$m PAH features fitting Gaussian
profiles to the lines and lines to the local continuum.
Our measurements of the [Ne\,{\sc ii}] flux for clusters M4, M5, M6
are in good agreement with those reported by \cite{Galliano2008} from
ground-based high angular resolution observations. 
Since the PAH features are
broad, it has been noted in the literature that Gaussian profiles might
not be appropriate to measure their flux because a large fraction of
the energy in these bands in radiated in the wings. A Lorentzian
profile might be a 
better approximation \citep[see][]{Galliano2008PAH} to measure their flux, and therefore we
repeated the line fits with this method. Table~\ref{tab:spectroscopy} lists 
the measurements for the extracted spectra for the Gaussian
  profiles. To illustrate the effect of using different profiles, in
  this table we give the measured 6.2 to $11.3\,\mu$m PAH
  ratio for the two profiles and the selected regions.

\begin{table*} 
\centering
 \begin{minipage}{150mm}
  \caption{Measurements from the Spitzer/IRS SL1+SL2 spectra}\label{tab:spectroscopy}
  \begin{tabular}{lccccccccc}
  \hline
  Region    & Extraction  & \multicolumn{2}{c}{$6.2\,\mu$m PAH feature} &
\multicolumn{2}{c}{$11.3\,\mu$m PAH feature} & [Ne\,{\sc ii}]$12.81\,\mu$m & $S_{\rm
  Si}$ & \multicolumn{2}{c}{$\frac{{\rm PAH}6.2}{{\rm PAH}11.3}$}\\
            & Aperture & flux           & EW & flux & EW & flux & & G & L\\
\hline
AGN & $3.7\arcsec \times 3.7\arcsec$ & 8.6 & 0.09 & 5.3 & 0.13 & 1.7 &
$-0.10$ & 1.6 & 1.2\\
M2+M3 & $5.6\arcsec \times 3.7\arcsec$ & 21.0 & 0.50 & 12.7 & 0.61 &
4.6 & $-0.24$ & 1.7 & 1.1\\
M4    & $3.7\arcsec \times 3.7\arcsec$ &10.7 & 0.45 & 5.8 & 0.63  & 2.3 &
$-0.85$ & 1.8 & 1.7\\
M5+M6 & $5.6\arcsec \times 3.7\arcsec$ & 20.0 & 0.45 & 10.2 & 0.40 &
6.5 & $-0.81$ & 2.0 & 1.5\\
Integrated & $27.8\arcsec \times 24.0\arcsec$ & 265.0 & 0.42 & 161.0 &
0.49 & 60.5 & $-0.45$ & 1.6 & 1.3\\ 
\hline

\end{tabular}

Notes.--- Fluxes (observed, not corrected for extinction) are in units of 
$\times 10^{-13}\,{\rm erg \, cm}^{-2}\,{\rm s}^{-1}$ and EW in units
of $\mu$m for measurements done with Gaussian profiles.   The
  typical errors are 10\% for the fluxes and $0.05\,\mu$m for the
  EW. $S_{\rm Si}$ is the strength of the silicate feature (see Section~\ref{sec:spectralmaps} for the
definition). The ratio of the
$6.2\,\mu$m PAH flux to
the $11.3\,\mu$m PAH feature flux is given for fits to the features
done with
Gaussians (G) and Lorentzian (L) profiles (see
Section~\ref{sec:1Dspectra} for more details).  
\end{minipage}
\end{table*}


\begin{figure}
\resizebox{1.\hsize}{!}{\rotatebox[]{-90}{\includegraphics{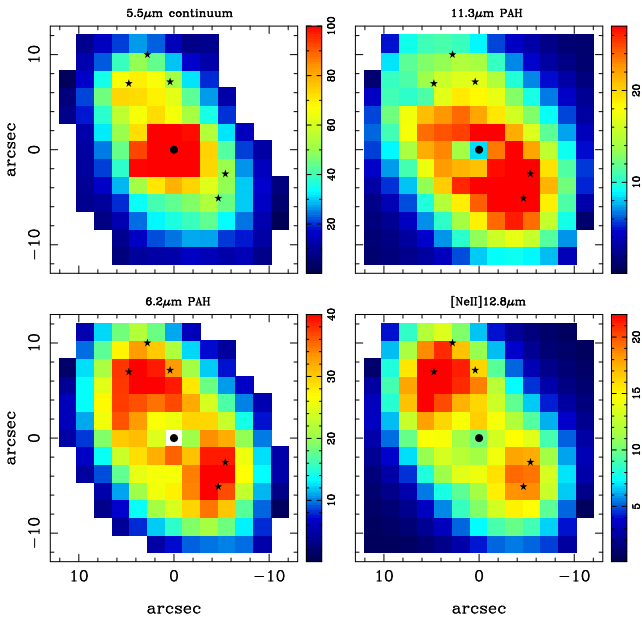}}}
     \caption[]{Spitzer/IRS observed (not corrected for extinction)
       spectral maps 
built with {\sc
         cubism} from the SL data cubes.  The maps of the  
$6.2\,\mu$m PAH feature  (lower left), $11.3\,\mu$m PAH feature
(upper right),   and [Ne\,{\sc
  ii}]$12.81\,\mu$m (lower right) are in units of $ 10^{-7}\,{\rm
  W}\,{\rm m}^{-2}\,{\rm sr}^{-1}$, and the $5.5\,\mu$m continuum map
(upper left) in units of ${\rm MJy}\,{\rm sr}^{-1}$. 
Only pixels with relative errors of less than 20\%
       are displayed. The position of the AGN (filled dot) was made to
       coincide with the peak of the $5.5\,\mu$m continuum
       emission. The positions of the 
M2...M6 mid-IR
       clusters (filled star symbols) of \cite{Galliano2005} 
are also plotted. Orientation
is north up, east  to the left. The apparently faint $11.3\,\mu$m
PAH  emission in the region of 
clusters M4, M5, and M6 is due to foreground
  extinction (see also the left panel of Fig.~7, and
  Sections~\ref{sec:morphology_detailed} and \ref{sec:silicatefeature}).}   
        \label{fig:spectralmaps}
    \end{figure}

\subsubsection{Spectral Maps}\label{sec:spectralmaps}
We used  {\sc cubism} to construct spectral maps of the most prominent
features in the SL data cubes, namely, the 
$6.2\,\mu$m and $11.3\,\mu$m PAH features, and
the [Ne\,{\sc ii}]$12.81\,\mu$m fine structure line.   The technique 
used here was very similar to that of \cite{AAH09} 
and involves  integrating  the line flux over 
the user-defined emission line  regions.  
Note that, unlike the line measurements in the previous section, the features are
not actually fitted with Gaussian or Lorentzian profiles, and
therefore these maps are only used for morphological purposes.
Since {\sc cubism} does not fit or deblend  emission
lines, the SL  
[Ne\,{\sc ii}]$12.81\,\mu$m map includes some contribution from the
$12.7\,\mu$m  PAH feature.
We also built a continuum map at $5.5\,\mu$m.
The Spitzer/IRS SL spectral maps shown in Fig.~\ref{fig:spectralmaps}
were  trimmed and rotated to the usual orientation 
of north  up, east to the left. We used the associated uncertainty
maps produced by {\sc cubism} to compute the relative errors of the
spectral maps. 

\begin{figure}
\resizebox{1.\hsize}{!}{\rotatebox[]{-90}{\includegraphics{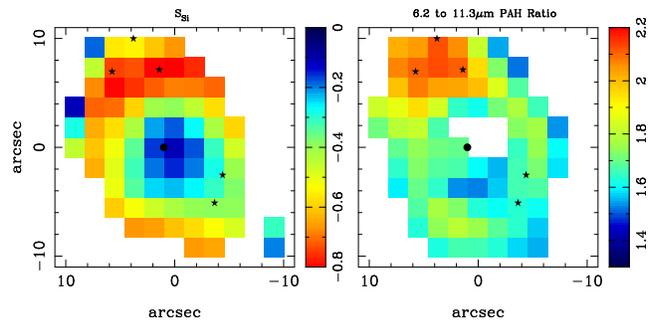}}}
\vspace{-2.5cm}        

     \caption[]{Spitzer/IRS spectral maps of the strength of the
       silicate feature 
       $S_{\rm Si}$ (left panel) and the observed (not corrected for
       extinction) ratio of the 
$6.2\,\mu$m to the $11.3\,\mu$m PAH feature emission
(right panel). Negative values of $S_{\rm Si}$ indicate that the
feature is observed in absorption. 
Orientation is north up, east to the left. Symbols are
as in Fig.~\ref{fig:spectralmaps}.}   
        \label{fig:silicatesPAHratio}
    \end{figure}

Finally we constructed the map of the silicate feature, which is shown in
Fig.~\ref{fig:silicatesPAHratio}, following the technique of 
\cite{Pereira2010IRSmapping}. Briefly, it involves fitting the silicate
feature from 1D spectra extracted in $2{\rm pixel} \times 2{\rm
  pixel}$ boxes from the IRS SL data cubes. The map is then
constructed by moving by 1 pixel in the x and y directions until the
FoV of the IRS SL data cubes is completely covered. 
We measured the apparent strength of the silicate feature in the IRS
spectra  following \cite{Spoon2007}:
$S_{\rm Si} = \ln f_\nu({\rm obs})/f_\nu({\rm cont})$, where
$f_\nu({\rm obs})$ is the flux density at the feature and $f_\nu({\rm 
  cont})$ is the flux density of the underlying continuum. The latter was 
fitted as a power law between
5.5 and $14\,\mu$m. We evaluated the strength of the silicate feature
at $10\,\mu$m. When the silicate strength is negative, the 
silicate feature is in absorption, whereas a positive silicate
strength indicates that the feature is seen in emission. 
The uncertainties of the measured strengths in the spectral map and
the extracted 1D spectra of previous section are
$\pm 0.05$.

\section{AGN Infrared Emission}\label{sec:spectraldecomposition}

A number of works have studied in detail the mid-IR emission of the
AGN hosted by 
NGC~1365 using high angular resolution observations, including
imaging, spectroscopy, and interferometry. Our T-ReCS nuclear ($\sim
0.35\arcsec$) mid-IR spectrum of NGC~1365  (see Fig.~\ref{fig:torusmodel})
is an almost featureless continuum with no clear evidence of the
presence of the
$9.7\,\mu$m silicate feature or PAH features 
\citep[see also][]{Siebenmorgen2004, Tristram2009}. By 
contrast,  the spectrum integrated over several arcseconds  is rich
in spectral features related to  star formation activity, such as, PAH
features and the [Ne\,{\sc ii}] line
(see Fig.~\ref{fig:SLspectra}).  
\cite{Tristram2009} modelled mid-IR
interferometric observations of the nuclear region and derived a
size (diameter) of $1.1-2.7\,$pc for the $12\,\mu$m emitter.
Up until now, there has been no estimates of the far-IR emission
arising from the AGN of NGC~1365, and thus the main goal of this
section is to provide such an estimate.

On scales of a few arcseconds the AGN of NGC~1365 is surrounded by a number of
mid-IR bright clusters and extended and diffuse emission 
(see Figs.~\ref{fig:IRcentralimages} and \ref{fig:closeups}, and
Section~\ref{sec:spatiallyresolved}). Moreover, in 
the far-IR AGN emission becomes  less luminous for increasing
wavelengths when compared to that of the clusters
(see Fig.~\ref{fig:IRcentralimages}). 
This together with the limited angular resolution of the PACS images
(i.e., the best angular resolution is FWHM$\sim 5.6\arcsec$ at $70\,\mu$m)
prevents us from doing aperture photometry on the PACS images to derive the 
far-IR flux densities of the AGN. Instead, in this section we
estimate the AGN emission by both fitting the nuclear near- and mid-IR
SED with torus models and doing a spectral decomposition of the
Spitzer/IRS nuclear spectrum.

\begin{table}
\centering
 \begin{minipage}{80mm}

\caption{Parameters of the \textit{CLUMPY} Torus Models}\label{tab:torusmodels}
\begin{tabular}{ll}
\hline
Parameter & Interval\\
\hline
Torus radial thickness  $Y$ & [5, 100]\\
Torus angular width   $\sigma_{\rm torus}$ (deg) & [15, 70]\\
Number clouds along an equatorial ray  $N_0$ & [1, 15]\\
Index of the radial density profile  $q$ & [0, 3]\\
Viewing angle  $i$ (deg) & [0, 90]\\
Optical depth per single cloud  $\tau_V$ &  [5, 150]\\
\hline
\end{tabular}
\end{minipage}
\end{table}
\bigskip

\subsection{Fit to the nuclear SED and mid-IR spectrum 
using clumpy torus models}\label{sec:torusfit}

 Clumpy torus models \citep[e.g.,][]{Honig2006,
Schartmann2008} and in particular the \cite{Nenkova2008a, Nenkova2008b}
models (also known as {\sc clumpy} models) 
have been shown to fit well the near and mid-IR SEDs and spectroscopy
of Seyfert galaxies  
\citep{RamosAlmeida2009,
  RamosAlmeida2011AGN, Mason2009, Nikutta2009, 
AAH11AGN, Sales2011, Lira2012}.  The {\sc clumpy} 
torus models are described by six
parameters to account for the torus geometry and the properties of the
dusty clouds  (see Table~\ref{tab:torusmodels}). The geometry of the torus 
is defined with the torus radial thickness $Y$, which is the 
ratio between the
outer and inner radii of the torus $Y = R_{\rm o}/R_{\rm d}$\footnote{The inner
  radius of the torus in 
these models is set by the dust sublimation temperature, which 
is assumed to be $T_{\rm sub} \sim 1500\,$K, and the AGN bolometric 
luminosity $L_{\rm bol} ({\rm AGN})$. Then the dust sublimation radius
in pc is $R_{\rm 
  d}=0.4\,L_{\rm bol} ({\rm AGN})^{0.5}$, where the AGN bolometric 
luminosity is in units of $10^{45}\,{\rm erg \,s}^{-1}$.}, 
and the width of the angular distribution  of the clouds $\sigma_{\rm
  torus}$, that
is, the angular size of the torus as measured from its equator.
The clouds have an optical depth $\tau_V$ and are arranged with a 
radial distribution expressed as a declining power law with index $q$
($\propto r^{−q}$), with $N_0$ giving the mean
number of clouds along a radial equatorial ray. 
The last parameter is the viewing angle to the  torus $i$. We refer
the reader to the sketch of the {\sc clumpy} model geometry in Figure~1
of \cite{Nenkova2008a}. 

\begin{figure}
\resizebox{1.\hsize}{!}{\rotatebox[]{-90}{\includegraphics{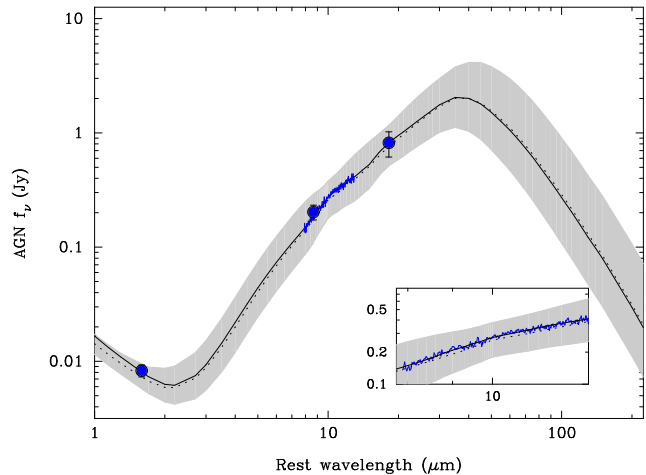}}}

\vspace{-1cm}
     \caption[]{Best fit to the nuclear SED (blue dots) and mid-IR
       spectroscopy (blue line) of 
NGC~1365 (MAP and median fits are the solid and dotted lines, respectively)
using the {\sc clumpy} torus models (including an AGN component). The shaded region  represents the  
range of acceptable models with a 68\% confidence level.  The inset
shows the spectral region around the $9.7\,\mu$m silicate feature. 
}  
        \label{fig:torusmodel}
    \end{figure}

We updated the fit of \cite{RamosAlmeida2011AGN}
with two main differences. First, we  used the higher
angular resolution mid-IR unresolved flux densities inferred in
Section~\ref{sec:trecs} and also included the T-ReCS mid-IR spectrum
for the fit. 
Second, this galaxy is classified as a Seyfert 1.5 \citep[it
shows a broad component in H$\beta$,][]{Schulz1999} and therefore there is
an unobscured view to the AGN. To account for it, we added an AGN component to
the resulting torus model (see below) and allowed for a small amount
of foreground 
extinction \citep[$A_V <5\,$mag, see][]{Alloin1981}, to redden the
fitted models. We also included in the
nuclear SED the HST/NICMOS $1.6\,\mu$m  unresolved flux \citep{Carollo2002}.

Similarly to our previous work \citep[][]{RamosAlmeida2011AGN,
  AAH11AGN}, we took 
a Bayesian approach to both fit the torus models to the data and derive
meaningful confidence levels for the fitted parameters. To this
end, we used the BayesClumpy (BC) fitting
routine, an interpolated version of the {\sc clumpy} models
\citep{AsensioRamos2009}. The new version of BC allows to fit both 
photometric points and spectra. For the  modelling 
we used the whole range of parameters probed by the {\sc clumpy}
models. We list them in 
Table~\ref{tab:torusmodels}. Unlike in some of our previous work, here
we allowed for the torus extent to vary in the range $Y=5-100$,
instead of restricting the models to small tori. This is
because larger tori may be required to  
reproduce the far-IR emission of  AGN
\citep{RamosAlmeida2011Herschel}.  
In addition to the six torus model parameters to be fitted plus the
foreground extinction, there
is an extra parameter to account for the vertical displacement
needed to match the fluxes of a given model to the observations.
This vertical shift, which we allow to vary freely, scales with
the AGN bolometric luminosity \citep[see][]{Nenkova2008b}.
We assumed uniform distributions in the ranges given in
  Table~\ref{tab:torusmodels} for 
the prior distributions of the torus model parameters.

The BC fitting routine translates the probability
distributions of the fitted torus model parameters
 into two model spectra. The first corresponds to the
maximum-a-posteriori (MAP) values that represent the best
fit to the data, while the second is the model produced with the
median value of the probability distribution of each parameter.
Fig.~\ref{fig:torusmodel} shows the  best fit to the nuclear
near and mid-IR SED and the mid-IR spectrum of the NGC~1365 AGN. We
also plot in  this figure the range of acceptable torus models  at the
$68\%$ confidence level. As can be seen from this figure, the fit to
the  photometric and spectroscopic data is very good. 
  Specifically, the {\sc clumpy} models reproduce well the 
flat silicate feature at around $9.7\,\mu$m
\citep[see][their figures 16 and 17]{Nenkova2008b}.

We can now compare the inferred 
torus model parameters with observations. We start with the AGN
  bolometric luminosity. As explained above, we chose to keep the
  scaling of the torus models to the observations as a free
  parameter. Because the scaling is proportional to the AGN bolometric
  luminosity, when compared with the observations it 
provides information about the goodness of the fit \citep[see][for
more details]{AAH11AGN}. The
  derived AGN bolometric luminosity from our fit is $L_{\rm bol}({\rm AGN}) = 2.6
  \pm 0.5 \times 
10^{43}\,{\rm erg \, s}^{-1}$. \citet{Risaliti2005} infered
a $2-10\,$keV luminosity in the range $0.8-1.4\times 10^{42}\,{\rm erg
  \, s}^{-1}$ (corrected for the assumed distance in this work). Our
derived AGN bolometric luminosity is consistent with the values from
X-rays after
applying a bolometric correction.

 We can also compare the  torus radius $R_{\rm
    o}=5^{+0.5}_{-1}\,$pc and  torus angular width $\sigma_{\rm
  torus}=36^{+14}_{-6}\deg$ from the modelling with observations.
The width of the angular 
distribution of the clouds in the torus is related to the
opening angle of the ionization cones, with some dependence with the
geometry of the gas \citep[see discussion in ][]{AAH11AGN}.
The half opening angle of the NGC~1365 ionization cone
modeled by \cite{Lindblad1999} is $\Theta_{\rm cone}=50$deg (measured
from the pole), which is 
compatible with the derived angular size of the torus. 
The size (diameter) of the $12\,\mu$m emitting source
inferred from the modelling of the mid-IR interferometric observations
of this galaxy is $s_{12\mu{\rm m}}=1.1-2.7\,$pc
\citep{Tristram2009}. We note that the size of the 
$12\,\mu$m emitting source, which traces the warm dust within the
torus, is expected to be smaller than
the {\it true} size of the torus. This is because the $12\,\mu$m 
emission is rather insensitive to cooler material further from the
nucleus \citep[see discussion in][]{AAH11AGN}. Given the good
agreement with the observations, we conclude that the fitted torus model
provides a good representation of the AGN torus emission.

\subsection{AGN torus emission in the far-IR}\label{sec:farIRAGNemission}
To estimate the AGN far-IR flux density in the
PACS bands, we extrapolated the fitted torus model 
beyond $\sim 20\,\mu$m. As can be seen from
Fig.~\ref{fig:torusmodel}, the  
torus model flux density peaks at $40-50\,\mu$m and falls off
rapidly at longer wavelengths. This is similar to results for other
AGN \citep[see e.g.,][]{Mullaney2011}. 
In Table~\ref{tab:AGNfluxes} we list 
the extrapolated AGN far-IR
flux densities together with those in the mid-IR  from the 
T-ReCS imaging and spectroscopy. The ranges given in this table for the 
far-IR  fluxes from the BC fit of the NGC~1365 take into account the
$\pm 1\sigma$ 
confidence region of the fitted {\sc clumpy} torus models. 

A quick
comparison between the extrapolated AGN far-IR fluxes and the aperture
photometry of the central regions (Table~\ref{tab:photometry}) 
clearly shows that the AGN contribution at these wavelengths is small
and decreases for 
increasing wavelengths. Within the central 5.4\,kpc ($r=30\arcsec$) the AGN
 is predicted to contribute $15\%$ ($+8\%$, $-6\%$) of the total
  $24\,\mu$m emission. At
$70\,\mu$m  the AGN accounts for at most  2\% of the emission in
the same region and less than 1\% at 100 and $160\,\mu$m. Using the
flux densities measured for $r=30\arcsec$ and  the
  following equation from  
\cite{Dale2002}, 

\begin{eqnarray}
L_{3-1000\mu{\rm m}} = 1.559\,\nu L_{\nu}(24\mu{\rm m}) +
0.7686\, \nu L_{\nu}(70\mu{\rm m}) + \nonumber \\
1.347 \,\nu L_{\nu}(160\mu{\rm m}) 
\end{eqnarray}

\noindent we estimate a $3-1000\,\mu$m luminosity within the
ILR region of $\sim 9
\times 10^{10}\,{\rm L}_\odot$. The ratio between the AGN bolometric luminosity
and the IR luminosity of this region is then $0.05\pm0.01$. This is in
good agreement with findings for  the majority of local LIRGs hosting an AGN
\citep{AAH11LIRGs}.

\section{Spatially-resolved IR emission in the ILR region}
\label{sec:spatiallyresolved}

\subsection{Morphology}\label{sec:morphology_detailed}

The central region of NGC~1365 is a bright 
source of mid- and far-IR emission \citep[see e.g.,][]{Telesco1993,
  Galliano2005,Wiebe2009}, although its contribution
to the integrated emission of the galaxy decreases significantly at
longer wavelengths. 
The mid-IR morphology, especially in the IRAC 5.8
and $8\,\mu$m and in the T-ReCS images, 
resembles that of a circumnuclear ring of
star formation with a diameter of approximately 1.8\,kpc (see 
Figs.~\ref{fig:IRcentralimages} and \ref{fig:trecs}).  
The AGN and the bright mid-IR clusters M4,
M5, and M6 identified (see Table~\ref{tab:sources}) by \cite{Galliano2005} out to $\lambda \sim
12\,\mu$m are also clearly
identified in the Gemini/T-ReCS image at $18.3\,\mu$m  (see
Fig.~\ref{fig:trecs}). The region to 
the southwest of the AGN is more diffuse and similar to the H$\alpha$ 
morphology (see Fig.~\ref{fig:closeups}, upper panel). We do not,
however, identify
any bright compact sources there. This diffuse emission  is
 reminiscent of the optical morphology of the region containing
 superstar clusters SSC3 and SSC6 along with a number of less
 luminous optical clusters \citep[][]{Kristen1997,Lindblad1999}.


In the Spitzer/MIPS $24\,\mu$m image the emission from the AGN is 
still differentiated from that of the mid-IR clusters, although the
latter are not resolved individually because of the decreased angular
resolution. 
The Herschel observations of NGC~1365 show that   
the far-IR AGN emission becomes faint when compared to that of the clusters. 
This is not only because the clusters in the
central region of NGC~1365 are still intrinsically bright in the
far-IR, but also because in general 
the AGN emission is observed to fall 
steeply in the far-IR \citep[see e.g.,][and references
therein]{Netzer2007, Mullaney2011}. Moreover,  in general the far-IR emission of
AGN appears to be 
dominated by a starburst
component \citep{Hatziminaoglou2010}.
It is also apparent from Fig.~\ref{fig:IRcentralimages} that the group
of clusters M5...M8 are brighter than the group of M2 and M3 in the
far-IR, as is also the case in the mid-IR. This implies, as will be
discussed further in Section~\ref{sec:sfr}, that the obscured star
formation rate (SFR) 
is higher in the former group than in the latter. Moreover, the 
morphology of the SFR tracers (e.g., the $8$, $24$, and $70\,\mu$m
emissions) is similar to that of the 
molecular gas, especially that of the rare C$^{18}$O(J=2-1)
isotope \citep{Sakamoto2007}, where a ``twin-peak'' morphology is
observed. Such molecular gas twin peaks are observed in barred
galaxies \citep{Kenney1992} 
and are formed where dust lanes intersect nuclear rings of
star formation, as is the case of NGC~1365.

We can also compare the IR
morphology, in particular at 8, 24, and $70\,\mu$m, of the ILR
region with that of 
H$\alpha$\footnote{Note that the H$\alpha$+[N\,{\sc ii}] image
shown in Fig.~\ref{fig:closeups} has not been continuum subtracted,
and thus we restrict our discussion to the H$\alpha$ bright hot spots
identified by other works
\citep[e.g.,][]{Alloin1981,Kristen1997}.}. Since 
all these emissions probe the SFR in galaxies
\citep{Kennicutt1998,Kennicutt2009,AAH06,Calzetti2007,Rieke2009,Li2010}, a good
morphological correspondence on scales of hundreds of parsecs 
is expected. Fig.~\ref{fig:closeups} indeed shows that the H$\alpha$
hot spots in the ILR region \citep[Table~4, and also][]{Alloin1981,Kristen1997}  are clearly
associated with IR emitting regions (compare with the IRAC $8\,\mu$m
image as it has the best angular resolution). This seems to be the
case out to $\lambda=100\,\mu$m. At longer wavelengths the
decreased angular resolution of the Herschel images does not allow us to
 compare the morphologies. The opposite is not
necessary true because the H$\alpha$ emission is strongly affected 
by extinction caused by the prominent dust lane crossing the ILR region. Mid-IR
clusters M4 and M5 appear to be detected in the H$\alpha$ image, but
they are very faint. This is because they are in one of the regions
with the highest extinction, as we shall see in more detail in
Section~\ref{sec:silicatefeature}.

The Spitzer/IRS map of the [Ne\,{\sc ii}]$12.81\,\mu$m emission
covering the central 2.2\,kpc of NGC~1365 is shown in
Fig.\ref{fig:spectralmaps}. This emission line is a good tracer of the
current SFR because its luminosity is proportional to the number of
ionizing photons \citep{Roche1991, HoKeto2007}.
The [Ne\,{\sc ii}] emission appears as a partial ring and it is
  enhanced in the mid-IR clusters. Its morphology is similar to the PACS
  $70\,\mu$m morphology and  the CO ``twin 
peaks'' discussed by \citet{Sakamoto2007}. 
The AGN of NGC~1365, on the other hand is not
a bright [Ne\,{\sc ii}]$12.81\,\mu$m source, as found for other
  active galaxies \citep[see e.g.,][]{PereiraSantaella2010}.
Analogously to other SFR
indicators, namely the 24 and $70\,\mu$m emissions, the region to the
northeast of the AGN (clusters M4, M5, and M6) is brighter in
[Ne\,{\sc ii}] than the
region of clusters M2 and M3. The regions containing these bright mid-IR 
clusters account for approximately $20-25\%$ of the total [Ne\,{\sc
  ii}]$12.81\,\mu$m emission in the central 2.2\,kpc (see
Table~\ref{tab:spectroscopy}). Note, however that the emission from
the clusters is unresolved at the Spitzer/IRS SL angular resolution
and we did not apply a correction for unresolved emission. Therefore 
this contribution from the clusters should be taken as a lower
limit.

Figure~\ref{fig:spectralmaps} also shows the maps of the 6.2 and the
$11.3\,\mu$m PAH features. Both features probe mostly the 
emission from B stars rather than that  from  massive on-going star
formation \citep[O stars, see][]{Peeters2004}. The $6.2\,\mu$m PAH morphology 
is very similar to [Ne\,{\sc ii}], whereas the $11.3\,\mu$m
PAH map appears to have a deficit of
emission in the region to the northeast of the AGN. The map of the ratio
between these two features is in
Fig.~\ref{fig:silicatesPAHratio}. Although several works 
found variations in the PAH ratios of galaxies not explained by changes in
the extinction \citep[see e.g.,][]{Galliano2008,
  Pereira2010IRSmapping}, in the case of the central region of 
NGC~1365 it might be entirely due to
extinction. The 6.2 to $11.3\,\mu$m PAH ratio is strongly increased in
the region enclosing clusters
M4, M5, M6, which is also the region suffering the highest extinction (see
Section~\ref{sec:silicatefeature} and Fig.~\ref{fig:silicatesPAHratio}).
Since the relative absorption at $11.3\,\mu$m is higher than at 
$6.2\,\mu$m \citep[see e.g., table~5 in][]{Brandl2006}, 
 correcting the $11.3\,\mu$m PAH map for extinction would
produce a morphology similar to that of the $6.2\,\mu$m PAH map. 
The measured EW and PAH ratios of the regions not affected by
  strong extinction are similar to those measured in
  other starburst galaxies \citep[see e.g.,][]{Brandl2006, Spoon2007, 
Sales2010}. 
We conclude that the star formation activity inside the 
 ILR region of NGC~1365 has been taking place for at least a few tens of
million years, as the PAH features trace the emission from B stars
\citep{Peeters2004}, and is taken place currently because of the
bright [Ne\,{\sc ii}]$12.81\,\mu$m emission requires the presence of
young ionizing stars.

\subsection{Extinction: Silicate Feature}\label{sec:silicatefeature}
The circumnuclear region of NGC~1365 is  crossed by a prominent dust
lane entering the ring \citep[see][and references
therein and also Figs.~\ref{fig:PACS100_large} 
and \ref{fig:closeups}]{Lindblad1999}.  This dust lane is still apparent in the
shortest wavelength IRAC bands and can be seen passing between
clusters M5 and M6 at $3.6\,\mu$m (see Fig.~\ref{fig:IRcentralimages}).
The brightest mid-IR clusters M4, M5, and M6 are located
at the edge of the dust features where the bar dust lane enters the
 ILR  region to the
northeast of the AGN \citep[][and also Figs.~\ref{fig:IRcentralimages}
and \ref{fig:closeups}]{Elmegreen2009}.  Mid-IR clusters M2 and 
M3 to the southwest of the nucleus are
in a region less affected by extinction and are likely to be  associated with
optical super star clusters and optically-detected H\,{\sc ii} regions
\citep[][]{Kristen1997,
  Sakamoto2007}. 

\begin{figure}

\hspace{0.5cm}
\resizebox{0.9\hsize}{!}{\rotatebox[]{0}{\includegraphics{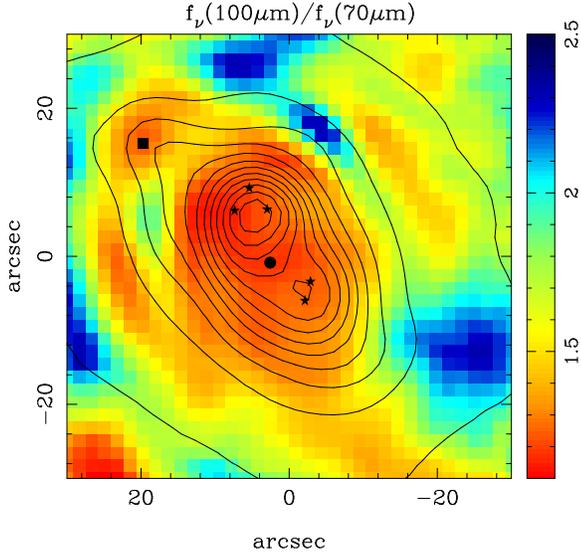}}}

     \caption[]{The color map is the Herschel/PACS
       $100\,\mu$m to Herschel/PACS $70\,\mu$m ratio. The displayed
       range corresponds to dust temperatures of approximately 
between 22\,K (high value
       of the ratio) and 32\,K (low value of the ratio). The contours are
       those of the PACS $100\,\mu$m image in a square root intensity 
scale as in 
Fig.~\ref{fig:IRcentralimages}. The FoV and symbols are also as in
Fig.~\ref{fig:IRcentralimages}. }  
        \label{fig:colortemperature}
    \end{figure}

The broad feature  at $\sim 10\,\mu$m is believed to be
produced by amorphous silicate grains
and in {\it normal} star-forming galaxies is observed in mild
absorption \citep{Roche1991, Smith2007SINGS}. In the simplest dust geometry of a
purely absorbing foreground screen the observed apparent depth of this
feature is proportional to the optical extinction. 
The general variation of the obscuration inside the
ILR region of NGC~1365 revealed by the optical imaging 
can be traced with the spectral map of the silicate
feature. As can be seen from  Fig.~\ref{fig:silicatesPAHratio}, 
the apparent strength of the silicate feature is high (in absorption)
in the region encompassing 
clusters M4, M5, and M6 to the northeast of the AGN, almost zero at the
location of the AGN, and intermediate (also in absorption) 
around clusters M2 and M3.

In Table~\ref{tab:spectroscopy} we list
the strength of the silicate feature
measured in the SL 1D spectra of selected regions. 
Assuming a foreground screen of dust and adopting the  \cite{Rieke1985} 
extinction law ($A_V/S_{\rm Si} = 16.6$), 
the observed apparent depths of the star-forming regions in the
circumnuclear region of NGC~1365 imply values
of the visual extinction of between $A_V \sim 4\pm1\,$mag and $A_V \sim
14\pm1\,$mag for the regions of clusters M2+M3,and M4+M5+M6,
respectively. The high value of the extinction derived for the latter region
is entirely consistent with the values
derived by \cite{Galliano2008} for the individual clusters using
hydrogen recombination lines. Our value of the extinction in the region
containing M2 and M3 is higher than the optical estimate for the L2 and L3
H\,{\sc ii} regions from \cite{Kristen1997}. 

The apparent strength of the silicate feature of the AGN measured from
the Spitzer/IRS SL spectrum is $S_{\rm
  Si}=-0.10\pm 0.05$ (Table~\ref{tab:spectroscopy})
indicating that the feature is present slightly in
absorption. This value, however, is  contaminated by extended  
emission as is apparent from the $8\,\mu$m image (see
Fig.~\ref{fig:closeups}) and the presence of PAH features in the
IRS nuclear spectrum (Fig.~\ref{fig:SLspectra}).  Indeed, the Gemini/T-ReCS 
high  angular resolution spectrum of the AGN is
mostly a featureless continuum with no evidence 
of PAH feature emission (Fig.~\ref{fig:torusmodel}). 
Finally, the relatively moderate strength of the
silicate feature of the integrated $\sim 30\arcsec \sim 2.7\,$kpc 
central region  ($S_{\rm
  Si}=-0.45\pm0.05$, Table~\ref{tab:spectroscopy}) of 
NGC~1365\footnote{The $8-1000\,\mu$m IR luminosity of this galaxy is
$\log (L_{\rm IR}/{\rm L}_\odot) = 11.03$ using the IRAS flux
densities from \cite{Sanders2003}.}
is typical of the observed
nuclear silicate  
strengths measured in other local LIRGs \citep{Pereira2010IRSmapping, 
AAH11LIRGs} and indicates an average visual extinction in this region 
of $A_V\sim 7\pm1\,$mag.

\subsection{Dust Color  Temperatures}\label{sec:colortemperature}
The Herschel images can be used to trace the spatial variations of 
the temperature of the dust 
inside the ILR region of NGC~1365. If we assume that the far-IR emission 
of a galaxy can be approximated with
a modified blackbody, then 
in the case of optically thin emission the flux density can be
expressed as

\begin{equation}\label{eq:modifiedblackbody}
f_{\nu} \propto \nu^{\beta}\,B(\nu, T_{\rm dust}) 
\end{equation}

\noindent where $B(\nu,T_{\rm dust})$ is the blackbody function for
a dust temperature of $T_{\rm dust}$ and $\beta$ is the
dust emissivity. In the simplest approximation we can calculate the 
color temperature $T_{\rm c}$ of the dust using the ratio of the 
surface brightness
at  two wavelengths with the following equation expressed in terms
of wavelengths

\begin{equation}\label{eq:colortemperature}
\frac{f_{\nu}(\lambda_1)}{f_{\nu}(\lambda_2)}=
\left(\frac{\lambda_2}{\lambda_1}\right)^{3+\beta}\,
\left(\frac{{\rm e}^{hc/\lambda_2KT_{\rm c}}-1}
{{\rm e}^{hc/\lambda_1KT_{\rm c}}-1}\right)
\end{equation}

We chose to construct a map of the PACS $100\,\mu$m to the PACS $70\,\mu$m
ratio as these two wavelengths provide the best angular resolutions
with Herschel. We
rebinned the PACS $70\,\mu$m image to the pixel size of the
PACS $100\,\mu$m image. We  matched the PSF 
of the two images by smoothing the $70\,\mu$m
with a Gaussian function, although this
may introduce some artifacts \citep[see e.g.,][]{Bendo2010,
  Bendo2012}. We then calculated  the dust color temperature by solving
Equation~\ref{eq:colortemperature}. We fixed the value of the dust 
emissivity $\beta=2$, as it
is found to fit well the integrated SEDs of local IR 
galaxies \citep{Dunne2001}. 

The map of the PACS $100\,\mu$m to $70\,\mu$m 
ratio of the inner $\sim 60\arcsec$  is
shown in Fig~\ref{fig:colortemperature}.
The observed range of $f_\nu(100\,\mu{\rm m})/f_\nu
(70\,\mu{\rm m})$ in the central region translates into values of the
dust color temperature of between  
$T_{\rm c} (100\mu{\rm m}/70\mu{\rm m})\sim 32\,$K and $T_{\rm c}(100\mu{\rm m}/70\mu{\rm
  m})\sim 22\,$K. Within the ILR region of NGC~1365
the highest  $T_{\rm c} (100\mu{\rm m}/70\mu{\rm m})$
color temperatures correspond to regions actively forming stars, that
is, the regions containing the mid-IR star clusters and
the L4 H$\alpha$ hot spot. These bright star forming regions
have color temperatures similar to, although slightly higher than, 
those of H\,{\sc ii} regions in
the spiral arms. The regions with  
the coldest  $T_{\rm c} (100\mu{\rm m}/70\mu{\rm m})$  in the ILR
region might be associated with some of the foreground dust features seen in the
optical images (see Fig.~\ref{fig:PACS100_large}). Finally, the region
of the AGN does not appear different in terms of the color temperature
when compared to the bright star forming regions. This is probably
because the AGN is faint in the far-IR (see next section) coupled with
the relatively coarse angular resolution of this map (i.e., that of
the PACS $100\,\mu$m).

\begin{figure}

\resizebox{1.\hsize}{!}{\rotatebox[]{0}{\includegraphics{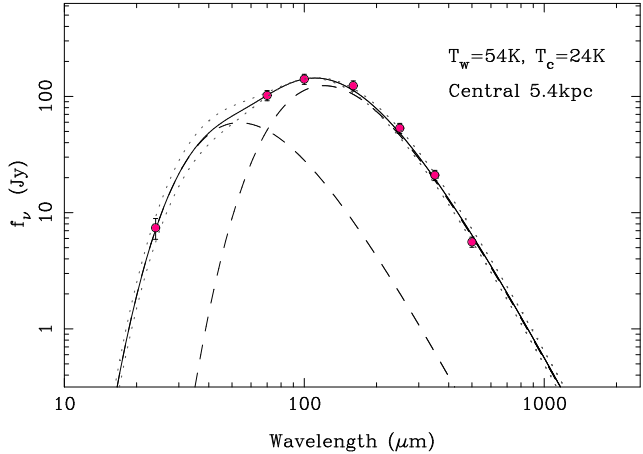}}}

     \caption[]{Best fit (solid line) to the AGN-subtracted 
        SED of the IR emission  in the ILR region of
       NGC~1365 ($r=30\arcsec$, filled dots). The best fit was achieved 
       using two modified blackbodies with temperatures $T_{\rm
         c}=24\,$K and $T_{\rm w}=54\,$K (dashed lines). The dotted
       lines are fits done by varying $T_{\rm c}$ by $\pm 1\,$K, and
       keeping the temperature of the warm dust fixed.}  
        \label{fig:SEDfit}
    \end{figure}

\subsection{Dust Properties}\label{sec:dustproperties}
In this section we study the properties of the dust in the  ILR region
of NGC~1365, particularly the dust heated by processes other than
the AGN. To do so, we first subtracted the AGN emission (see
Table~\ref{tab:AGNfluxes}) from the observed fluxes in the 
Spitzer/MIPS $24\,\mu$m band
and all the Herschel bands.  We then fitted
the observed non-AGN mid- and far-IR SED using a combination of two modified
blackbody following \cite{Dunne2001} and \cite{Clements2010}:

\begin{equation}
f_\nu = N_{\rm w} \nu^\beta B(\nu, T_{\rm w}) 
 + N_{\rm c} \nu^\beta B(\nu, T_{\rm c}) 
\end{equation}

\noindent where $N_{\rm w}$ and $N_{\rm c}$ are 
the relative masses of the warm and cold dust components and $T_{\rm
  c}$ and $T_{\rm w}$ are their temperatures. As done in
Section~\ref{sec:colortemperature}, we fixed the dust emissivity $\beta = 2$ and then used 
a standard $\chi^2$ minimization 
technique to fit the two dust temperatures and
the relative masses of the two dust components. For the fit we 
used the fluxes measured through an
$r=30\arcsec$ aperture, and normalized them
to that at $100\,\mu$m. For the Spitzer/MIPS $24\,\mu$m flux density, we added in quadrature the
photometric error and the 
uncertainty associated with subtracting the AGN component at this
wavelength to a total error budget of $\sim 20\%$. For the Herschel
data points because the AGN emission in the far-IR is very small
compared to the total emission inside the ILR region, the dominant source of
error is that associated with the photometric calibration of the data:
$\sim 10\%$.

Fig.~\ref{fig:SEDfit} shows the  best fit to the AGN-subtracted SED of
the ILR region of NGC~1365, which was obtained with $T_{\rm
         c}=24\,$K and $T_{\rm w}=54\,$K, and relative masses of cold
       to warm dust of $N_{\rm c}/N_{\rm w}\sim 120$. It is worth noting
       that the fit using two modified 
blackbodies is formally better than a fit to the far-IR data (i.e.,
excluding the $24\,\mu$m data point) using a single modified
blackbody.  This can be explained because there
is a non-negligible contribution from the warm dust component to the
Herschel PACS $70\,\mu$m flux density, 
as found for the integrated emission of other nearby galaxies
\citep[see e.g.,][]{Bendo2010, Smith2010}.
 However, the temperature of the warm
component is not tightly constrained since varying the warm dust
temperature by as much as $\pm 4\,$K produces statistically similar
good fits ($\chi^2 \le 2\,\chi^2_{\rm min}$). This is because the peak of
this component at around 
$50\,\mu$m is not well sampled with the present data and the mass contribution of the warm
component is small. The temperature of the cold component,
on the other hand is well constrained. To get an estimate of the
uncertainties on the fitted cold dust temperature, we changed it while fixing the temperature of the warm component. As can
be seen from Fig.~\ref{fig:SEDfit}, cold dust temperatures in the
range $T_{\rm c}=24\pm 1\,$K 
provide acceptable fits to the SED.

Previous works showed that the integrated mid-IR and far-IR SEDs of Seyfert
galaxies could be fitted with a combination of various dust
temperatures probing different heating mechanisms. These include
warm dust associated with the AGN torus,
cold dust similar to that observed in starburst galaxies, and very cold
dust at temperatures characteristic of dust heated by the interstellar
medium \citep{PerezGarcia2001,Spinoglio2002}. With the superior
angular resolution of the Herschel observatory we can now obtain
spatially resolved observations of nearby Seyfert 
galaxies and star forming galaxies. The dust temperature of the ring of
star formation in NGC~1365 is similar to that of other nuclear and circumnuclear
starbursts with or without an AGN with similar IR luminosities and
physical sizes, for instance, the ring of NGC~3081
\citep{RamosAlmeida2011Herschel} and the nuclear region of M83
\citep{Foyle2012}.   
On the other hand, the dust of Mrk~938, which is part of our survey of
Seyfert galaxies, has a considerably higher dust 
temperature of $T=36\pm 4\,$K, probably due to the smaller size of the
IR emitting region of this galaxy and higher IR luminosity 
\citep{Esquej2011}. 

We calculated the dust mass ($M_{\rm dust}$) within the  ILR region of NGC~1365 
using the following equation from \cite{Clements2010}
\citep[adapted from][]{Hildebrand1983}:

\begin{equation}
M_{\rm dust} = \frac{f_\nu\,D^2}{\kappa_{\rm dust}(\nu)} \times
\left( \frac{N_{\rm c}}{B(\nu,T_{\rm c})} + 
\frac{N_{\rm w}}{B(\nu,T_{\rm w})}\right)
\end{equation}

\noindent where $f_\nu$ is the observed flux density, $D$ is the
luminosity distance, 
$B(\nu,T)$ is the blackbody emission for the best fitting dust
temperatures, and $\kappa_{\rm d}(\nu)$ is the dust 
absorption coefficient. As done by \cite{Esquej2011}, we evaluated this
expression at $250\,\mu$m using an absorption coefficient of
$\kappa_{\rm dust}(250\mu{\rm m}) = 4.99\,{\rm cm}^2\,{\rm g}^{-1}$,  
as interpolated from the dust model of \cite{Li2001}. We derived a
dust mass in the ILR region of $M_{\rm dust}({\rm ILR})=6.9\times 10^7\,{\rm
  M}_\odot$, which accounts for approximately 25\% of the total dust mass 
of this galaxy $M_{\rm dust} ({\rm total}) = 3\times10^8\,{\rm
  M}_\odot$. The latter estimate is from \cite{Wiebe2009} but
recalculated for the absorption coefficient used in this work.

\subsection{Star Formation Rate}\label{sec:sfr}
The ages and masses of the mid-IR clusters \citep[$\sim 6-8\,$Myr
  and $\sim 10^7\,{\rm M}_\odot$, ][]{Galliano2008} and the large amount
  of molecular gas available in the central regions of this galaxy
  \citep[][]{Sakamoto2007} indicate that there is intense on-going
  star formation activity there.
We can use the IR 
observations of NGC~1365 to estimate the obscured SFR in the ILR
region, and compare it with the unobscured SFR. 
As discussed by
\cite{Kennicutt2009}, a combination of the observed (not corrected for
extinction) H$\alpha$ luminosity and a mid-IR monochromatic luminosity
(preferably $24\,\mu$m) will provide the best estimate of the total SFR in a
moderately obscured environment (see
Section~\ref{sec:silicatefeature}), such as the ILR region of 
NGC~1365. We note that this empirically calibrated recipe
  includes contributions from dust heating from all stars and not only
  the youngest \citep[see][for a full
  discussion]{Kennicutt2009}. Using the H$\alpha$ flux of \cite{Forster2004} 
 for a 40\arcsec-diameter aperture and our $24\,\mu$m flux (after
 subtracting the AGN component) we estimated a total SFR within the
 ILR region (inner $\sim 5.4\,$kpc, $r=30\arcsec$) of NGC~1365 of SFR$=7.3\,{\rm
   M}_\odot\,{\rm yr}^{-1}$ for a Salpeter initial mass function
 (IMF). Approximately 85\%
 of this SFR is contributed by the $24\,\mu$m emission, and thus 
it originates from dust-obscured star forming regions. 


Most of the IR emission in the ILR region comes from the star formation ring 
containing the bright mid-IR clusters identified by
 \cite{Galliano2005}. Indeed, we estimate that  
within the inner $2.7\,{\rm kpc}$ the total SFR is $5.6\,{\rm M}_\odot\,{\rm
  yr}^{-1}$. This has been calculated 
using the AGN-subtracted $24\,\mu$m flux density ($r=15\arcsec$) and the
24\arcsec-diameter  aperture 
H$\alpha$ flux from \cite{Forster2004}. This corresponds to a SFR
surface density 
in the  ring
of $\Sigma_{\rm SFR} =2.2 \,{\rm M}_\odot\,{\rm
  yr}^{-1}\,{\rm kpc}^{-2}$, for a ring radius of 900\,pc. This value
of the SFR density is similar to those of 
other circumnuclear starbursts \citep{Kennicutt1998} and 
is expected to be larger by factors of $100-1000$ compared to the disk of the
galaxy \citep{Elmegreen1994}. 

\section{Summary and Conclusions}

In this paper we have studied the IR emission associated with 
the star formation activity in
the ILR region of NGC~1365, as well as the IR emission of the AGN.
To this end we have analyzed new far-IR ($70-500\,\mu$m) Herschel/PACS
and SPIRE imaging, and high angular resolution ($\sim 0.4\arcsec$) 
Gemini/T-ReCS mid-IR imaging and spectroscopy of
this galaxy. We have also made use of  archival 
Spitzer/IRAC and MIPS imaging and IRS spectral mapping data.
Our main findings for the inner $D \sim 5\,$kpc
region of NGC~1365 are:
\begin{itemize}
\item
The new Herschel PACS imaging data at 70, 100, and $160\,\mu$m reveal that
the ring of star formation in the ILR region is bright in the
far-IR. The AGN is the brightest mid-IR source in the inner 2\,kpc up to
$\lambda\simeq 24\,\mu$m, but it becomes increasingly fainter in the far-IR
when compared with the mid-IR clusters or groups of them in the ring.
\item
The 24 and $70\,\mu$m emissions as well as the [Ne\,{\sc
  ii}]$12.81\,\mu$m 
line and PAH features trace
the star-forming ring in the ILR region and have morphologies similar to the
CO ``twin-peaks''. This all indicates that there is intense on-going 
star formation taking place in the inner few kpc of NGC~1365.
\item
The unresolved near and mid-IR nuclear emission and mid-IR spectrum (i.e., 
AGN-dominated emission) of NGC~1365 are well 
reproduced with a relatively compact torus (outer radius of $R_{\rm
  o}=5^{+0.5}_{-1}\,$pc) with an opening angle of $\sigma_{\rm torus}=36^{+14}_{-6}$deg, and an 
AGN bolometric luminosity $L_{\rm bol}({\rm
  AGN})=2.6\pm0.5\times 10^{43}\,{\rm erg \, s}^{-1}$
using the {\sc clumpy} torus models. These 
parameters are in good agreement
with independent estimates in the literature.
\item
Using the fitted torus model we quantified the AGN emission in the
far-IR. The AGN only contributes  at most 1\% of the $70\,\mu$m
emission within the inner 5.4\,kpc ($r=30\arcsec$), and less than
1\% at longer wavelengths. At $24\,\mu$m the AGN accounts for
$\sim 15\%$
of the emission in the same region. 
We estimated that the AGN bolometric contribution to
the $3-1000\,\mu$m luminosity in the inner 5.4\,kpc is approximately 5\%.
\item
The non-AGN 24 to $500\,\mu$m SED of the ILR region (inner 5.4\,kpc) of
NGC~1365 is well fitted with a combination of two modified blackbodies
with warm and cold temperatures of 54\,K and 24\,K,
respectively. However, the cold dust
component accounts for most of total dust mass inferred in this region
($M_{\rm dust}({\rm 
  ILR})= 7\times 10^7\,{\rm M}_\odot$)  and has a temperature similar
to that of other nuclear and circumnuclear starbursts of similar sizes
and IR luminosities.
\item
From the comparison between the SFR from H$\alpha$ (unobscured) and
the SFR from $24\,\mu$m (obscured) we infer that up to $\sim 85\%$ of the
on-going SFR inside the ILR region of NGC~1365 is 
taking place in dust-obscured regions in the ring of star formation.
\end{itemize}

\section*{Acknowledgments}

We are grateful to B. Elmegreen and E. Galliano for providing us with
the BVR map of NGC~1365 shown in the right panel of Figure~1. We also
thank Andr\'es Asensio Ramos for developing the BayesClumpy fitting
routine. We thank an anonymous referee for comments that helped
improve the paper.
 
A.A.-H., M.P.-S.,  and P.E. acknowledge support from the Spanish Plan Nacional
de Astronom\'{\i}a y Astrof\'{\i}sica under grant
AYA2009-05705-E. A.A.-H. also acknowledges support from 
the Universidad de Cantabria through 
the Augusto Gonz\'alez Linares Program and AYA2010-21161-C02-01. 
C.R.A. acknowledges the Spanish Ministry of Science and Innovation
(MICINN) through project Consolider-Ingenio 2010 Program grant
CSD2006-00070: First Science with the GTC
(http://www.iac.es/consolider-ingenio-gtc/) and the Spanish Plan
Nacional grant 
AYA2010-21887-C04.04.
M.P. acknowledges the Junta de Andaluc\'{\i}a and the 
Spanish Ministry of Science
and Innovation through projects PO8-TIC-03531 and AYA2010-15169, respectively.

{\it Herschel} is an ESA space observatory with science
  instruments provided by European-led Principal Investigator
  consortia and with important participation from NASA.
PACS has been developed by a consortium of institutes led by MPE
(Germany) and including UVIE (Austria); KU Leuven, CSL, IMEC
(Belgium); CEA, LAM (France); MPIA (Germany); INAF-IFSI/OAA/OAP/OAT,
LENS, SISSA (Italy); IAC (Spain). This development has been supported
by the funding agencies BMVIT (Austria), ESA-PRODEX (Belgium),
CEA/CNES (France), DLR (Germany), ASI/INAF (Italy), and CICYT/MCYT
(Spain).
SPIRE has been developed by a consortium of institutes led
by Cardiff Univ. (UK) and including: Univ. Lethbridge (Canada);
NAOC (China); CEA, LAM (France); IFSI, Univ. Padua (Italy);
IAC (Spain); Stockholm Observatory (Sweden); Imperial College
London, RAL, UCL-MSSL, UKATC, Univ. Sussex (UK); and Caltech,
JPL, NHSC, Univ. Colorado (USA). This development has been
supported by national funding agencies: CSA (Canada); NAOC
(China); CEA, CNES, CNRS (France); ASI (Italy); MCINN (Spain);
SNSB (Sweden); STFC, UKSA (UK); and NASA (USA). 

Based on observations obtained at the Gemini Observatory, which is
operated by the  Association of Universities for Research in
Astronomy, Inc., under a cooperative agreement 
with the NSF on behalf of the Gemini partnership: the 
National Science Foundation (United
States), the Science and Technology Facilities Council (United
Kingdom), the National Research Council (Canada), CONICYT (Chile), the
Australian Research Council 
(Australia), Minist\'erio da Ci\'encia, Tecnologia e Inova\c{c}\~ao (Brazil) 
and Ministerio de Ciencia, Tecnolog\'{\i}a e Innovaci\'on Productiva
(Argentina). 

This research has made use of the NASA/IPAC Extragalactic Database
(NED) which is operated by the Jet Propulsion Laboratory, California
Institute of Technology, under contract with the National Aeronautics
and Space Administration.

\label{lastpage}

\end{document}